**Possible Exciton Mechanism of Superconductivity in $Ag_5Pb_2O_6$/(CuO-$Cu_2O$) Composite**

Danijel Djurek

*Alessandro Volta Applied Ceramics (AVAC), Kesten brijeg 5, Remete, 10 000 Zagreb, Croatia*

**Abstract**

Composite material was prepared from Byström-Evers metallic particles $Ag_5Pb_2O_6$ dispersed in a semiconducting matrix CuO-$Cu_2O$. Partial substitution of CuO by $Cu_2O$ increases electric conductivity at 673 K for 2–3 orders of magnitude above Landauer conductivity $e^2/ha$ indicated by pure CuO matrix. Onset of superconductivity at 86–375 K presented in this and previous works is discussed in the framework of the exciton-polariton mechanism occurring on the grain surfaces.

**Introduction**

Byström-Evers (BE) metallic compound $Ag_5Pb_2O_6$ is a subvalent oxide with two Ag (1) cations bonded in a dimer (cluster) giving one valency to neighbour atoms [1]. Structure is trigonal (*P*31m) [2] as shown in Fig. 1, and $Ag_2$ dimers are positioned in c-axis channels. Dimensions of the unit cell are $a$ = 593.24, and $c$ = 641.05 pm, while Ag(1) – Ag(1) bond distance in dimer is $d$ = 309,30 pm. Byström and Evers pointed the high mobility of $Ag_2$ dimers along c-axis, while P. Pyykkö put forward [3] that their existence is a relativistic effect associated with $d^{10}$-$d^{10}$ orbital coupling. More detailed analysis of $d^{10}$–$d^{10}$ coupling was given by M. Jansen [4]. Pure $Ag_5Pb_2O_6$ is superconducting (SC) at $T_c$ = 52.4 mK [5].

Author and co-workers published several reports describing the occurrence of superconductivity in systems containing the $Ag_5Pb_2O_6$ [6]. More detailed analysis of data recorded from $Ag_5Pb_2O_6$ particles dispersed in various oxides and carbonates (PbO, CuO, $Pb_3O_4$, $PbO_2$, $PbCO_3$) appealed an attention for the possibility that reported SC events were not associated with bulk BE oxides, but with inter-grain properties [7], like point contact current transport or tunnelling.

It has been observed that CuO is the promising inter-grain material [8], which puts in doubt tunnelling effect as a single contribution to the conduction mechanism. In addition, x-ray diffraction pattern revealed small, but significant, fraction of $Cu_2O$ formed by reduction of CuO on the boundaries of BE particles at fusion temperature ~ 773 K. $Cu_2O$ is a classic exciton active semiconductor [9], and possible role of the surface excitons in the superconductivity has been put forward [10] by V. L. Ginzburg and D. A. Kirzhnits (GK). D. Allender, J. Bray and J. Bardeen (ABB) developed model [11] of an excitonic mechanism of superconductivity, and they assumed the tunnelling of metal electrons into the semiconductor gap. Electrons interact with virtual excitons which mediate electron-electron interaction, similarly as do phonons in known superconductors.

In further analysis, assumed inter-grain mechanism appealed an attention to the microcavity quantum electrodynamics developing rapidly in the past decade [12]. There are continuously increasing arguments in favour of the Bose-Einstein condensation, and possible room temperature superconductivity in the photon-exciton (polariton) continuum was proposed [13] by F.P. Laussy, A. V. Kavokin and I. A. Shelykh (LKS). In the reported experiments [14] photon field is amplified by multiple reflections on the cavity walls consisting of distributed Bragg reflectors, until amplification results in a Rabi splitting and formation of polaritons.

In this paper we report some clues which may be useful in analysis of the composite $Ag_5Pb_2O_6/(CuO–Cu_2O)$, and observed superconductivity may be possibly attributed to the exciton dynamics activated between the BE grain surface and semiconducting $CuO–Cu_2O$ matrix.

**Experiment**

Ag$_5$Pb$_2$O$_6$ was prepared from powders Ag$_2$O and PbO$_2$ (Alfa Aesar) by solid state reaction at 623 K and 150 bar O$_2$ [9]. The fused powder with particle size $a$ = 0.62 microns was then milled to the particle size $a$ = 0.25 microns. Independent mixing of CuO and Cu$_2$O was performed in a magnetic stirrer by use of ethanol as a mixing agent. Upon drying of the mixture, Ag$_5$Pb$_2$O$_6$ powder was added in the same way, and final system was labelled as (Ag$_5$Pb$_2$O$_6$)$_\alpha$ (CuO)$_\beta$ (Cu$_2$O)$_\gamma$, and weight proportions are given by $\alpha:\beta:\gamma$. Samples were prepared in the form as shown in the reference 7. Aluminium spacer of 1.5 mm thickness was used in order to ensure the same thickness of samples. Electric resistance was measured by use of four probe technique. In DC measurements Keithley model 6221/2182A combination was used, while lock-in amplifier SR 830 DSP was used for AC resistance measurements.

Resistive properties of CuO–Cu$_2$O system were investigated independently, and data for $\beta:\gamma$=1:0.2 are presented in Fig. 3. Samples were heated in an evacuated cell up to 773 K and cooled to the room temperature (RT). Increase of the resistance by cooling to RT indicates the semiconducting properties of CuO-Cu$_2$O system and increase of the conductivity with increasing fraction of Cu$_2$O sounds for the role of excitons.

Pellets were then prepared in the proportion $\alpha:\beta:\gamma$ = 1:1:0.2 and heated in an evacuated cell up to 773 K, maintained at this temperature 6–8 hours until BE oxide decomposes to Pb$_3$O$_4$ and Ag. Oxygen at 1.6 bar was then introduced in cell, and fusion of Ag$_5$Pb$_2$O$_6$ was re-established for 24 hour. Consumption of the oxygen was controlled by the capacitive pressure gauge, and manostat manifold was build, in order to maintain pressure at selected value during the solid state reaction. Pressure must be sufficiently high to induce the solid state reaction, and small enough to suppress chemisorption of the oxygen on grain surfaces and formation of surface states. Cooling to LN$_2$ temperature results in a resistive dependence as shown in Fig 4, and downturn of the resistance occurs at T$_C$ = 128,5 K. The resistivity measured at LN$_2$ temperature was less than 1.2·10$^{-8}$ Ωcm. In further experiments the fused samples were multiply reheated up to 620 K and cooled to RT in an evacuated chamber, which resulted in a gradual removal of the excess oxygen from the grain surface and increase of the surface resistance. Transition temperatures were lifted to higher values and an example is shown in Fig 5 after six heating cooling cycles, when final transition temperature reached 195 K for measuring AC current 1 mA. An application of the measuring AC current 50 mA re-established the resistive state.

At oxygen pressures > 5 bar applied during the fusion at 773 K adsorption is very efficient on the surface of BE particles, and $O_2$ up to ½ mole may be adhered, as evaluated by thermogravimetry and tensometry. The similar problem is oxygen chemisorption on the $Cu_2O$ grains [15] which is followed by the formation of the highly conducting layers. Chemisorption of the oxygen on the grain surfaces poses the most important technical problem in preparation of good samples. Multiple heating in an evacuated cell up to 673 K and subsequent cooling to RT favours the removal of excess oxygen.

**Discussion**

Partial substitution of CuO by $Cu_2O$ in $Ag_5Pb_2O_6$/CuO composite increases considerably the electric conductivity, as compared to the Landauer limit value $e^2/ha$ [16] measured in samples prepared by use of pure and semiconducting CuO. This sounds for the possible role of excitons in the conduction mechanism.

Dimerisation of Ag in $Ag_2$ clusters is relativistic effect known in the group of the coinage metals (Ag, Au, Pt), and relativistic contraction of Ag–Ag bond in AgH is 8 pm, while vibration frequency of $Ag_2$ dimer is $\nu = 7.7 \cdot 10^{12}$ Hz [17]. The latter corresponds to T = 364 K, and $Ag_2$ dimer may be considered as Einstein oscillator. Surface density of $Ag_2$ dimers is $\rho = 3.3 \cdot 10^{14} \cdot cm^{-2}$, while the surface density of polaritons is given by $n = \nu \rho P \tau$. $P \sim \exp(-E_0/kT_c)$ is probability of emission of the photon of energy $E_0$, and $\tau$ is lifetime of the exciton at T = 350 K. Available figure is $\tau = 5 \cdot 10^{-6}$ sec at T = 15 K [18]. Assuming this value, energy required to fit the surface density of polaritons $n = 7.5 \cdot 10^{11}$ cm$^{-2}$, as proposed by LKS at T = 350 K, is $E_0 = 0.71$ eV. Otherwise, calculation of the energy level difference for an electron on the grain surface in two dimensional hexagonal cell gives $E_0 = h^2/4ma^2 = 0.81$ eV. This corresponds to the surface electron density $\rho = 2\pi p_0^2/h^2 = 1/a^2 \sim 10^{15}$ cm$^{-2}$ given by GK, and assumed $a \sim 300$ pm, in good agreement with above value estimated for hexagonal BE lattice ($a/2 = 296.62$ pm). Finally, the relativistic contraction removes one valency, and it corresponds to the change of elastic energy $\Delta E \sim 1.85$ eV. Highly movable $Ag_2$ dimer, however, tends to escape from the grain boundary and stretch. This, in turn, results in valency fluctuation between two extreme values 1 and 2. This process is not energy conserving, and emission-absorption of electron or photon is a single option, despite of the comparatively low probability of such a process $P \sim 10^{-9}$.

Listed values should be compared to the quoted exciton band-gap energy in $Cu_2O$, $E_g$ = 2.172 eV at T =1.3 K [19] or $E_g$ = 2.01 eV at T = 350 K, obtained as an extrapolation of $E_g$ = 2.01 eV given T = 265 K [20]. My underestimate may also result from an application of the uniaxial stress which lifts the degeneracy at Brillouin zone centre ($\Gamma$ = 0) and reduces the band-gap energy, as it is the case in pure silicon [21].

Nevertheless, uncertainty in exciton lifetime $\tau$ affects little $E_0$ at T = 350 K, and polariton density calculated by LKS agrees with above estimated values within a half order of magnitude, which seems to be an encouraging result.

The cited approaches based upon exciton mechanism of superconductivity require either external photon source (GK and LKS) or virtual excitons created in the semiconducting layer on the metallic surface (ABB).

In this work there is assumed the possible emission of photons from the grain surface being in the contact with semiconducting matrix. Problem is posed by oxygen adsorbed on the grains, and this favours formation of the surface states and corresponding smearing of photon emission. $O_2$ must be removed, and this involves a serious technical problem which must be solved in the future work. Otherwise, partial cover of the grain surface by oxygen can serve as a tuning parameter for a polariton density. This probably explains why transition temperature is shifted to higher values after continuous heating–cooling cycles performed *in vacuo*. It is supposed that evacuation and cycling removes oxygen from the grain surface.

An alternative approach includes better preparation control by the formation of a sandwiched structure consisting of the thin layers $Ag_5Pb_2O_6$–semiconductor–$Ag_5Pb_2O$.

**Conclusion**

Composite material $Ag_5Pb_2O_6/(CuO$-$Cu_2O)$ was described as a possible candidate for exciton mediated superconductivity. It is worthy to note that V. L. Ginzburg proposed the metal granules imbedded in a dielectric matrix as a possible carrier of the excitonic superconductivity [22, 23]. The experimental data match several theoretical models dealing with the possibility of the ambient temperature superconductivity. Application of ABB model to $Ag_5Pb_2O_6/(CuO$-$Cu_2O)$ composite presumes tunnelling of electrons from the hexagonal surface lattice $Ag_5Pb_2O_6$ into the semiconductor $CuO$-$Cu_2O$ gap, while application of LKS model requires the persistent emission of photons from the grain surface, which excludes the necessity of an external laser excitation. A rather plausible ideas presented in this paper should be followed by more detailed calculation of surface states, while further experiments require the control of uniaxial stress and careful treatment of the grain surface, which includes the removal of adsorbed oxygen.

**Figure captions**

**Fig. 1** Unit cell of $Ag_5Pb_2O_6$. Red dots, connected with red lines, visualize the $Ag_2$ dimer.

**Fig. 2** c-axis view of $Ag_5Pb_2O_6$. Red dots, encircled by dashed circles, visualize $Ag_2$ dimers in hexagonal channels.

**Fig. 3** Temperature dependence of the resistance of $(CuO)_\beta (Cu_2O)_\gamma$ mixture, with $\beta:\gamma=1:0.2$.

**Fig. 4** Temperature dependence of the resistance of composite $(Ag_2Pb_2O_6)_\alpha (CuO)_\beta (Cu_2O)_\gamma$, with $\alpha:\beta:\gamma = 1:1:0.2$.

**Fig. 5** Temperature dependence of the resistance of composite $(Ag_2Pb_2O_6)_\alpha (CuO)_\beta (Cu_2O)_\gamma$, with $\alpha:\beta:\gamma = 1:1:0.2$ after six-fold heating up to 620 K and cooling to room temperature.

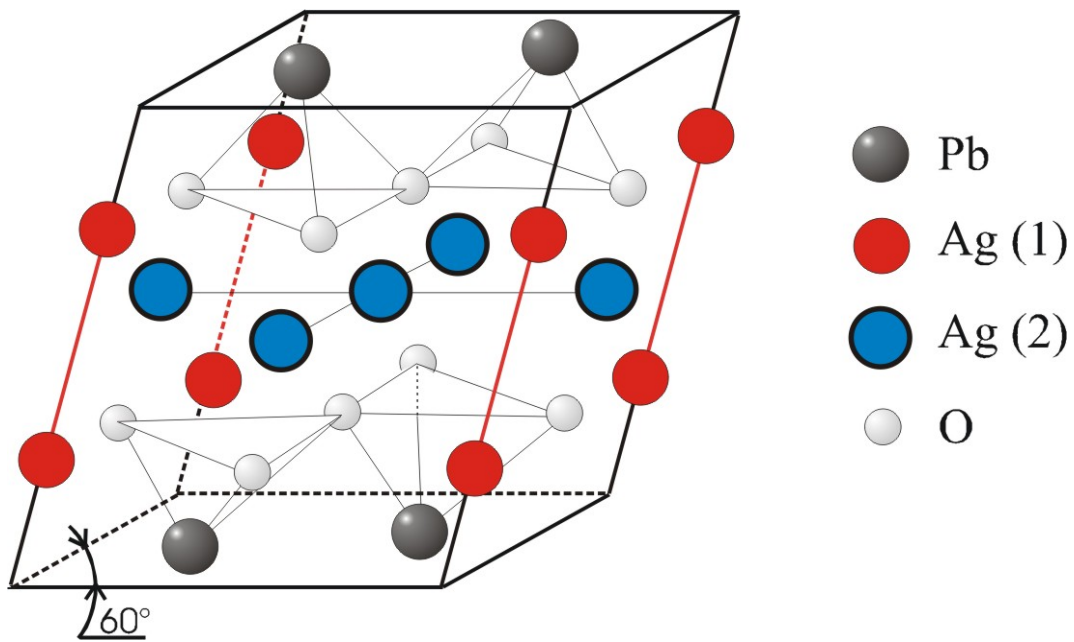

**Figure 1**

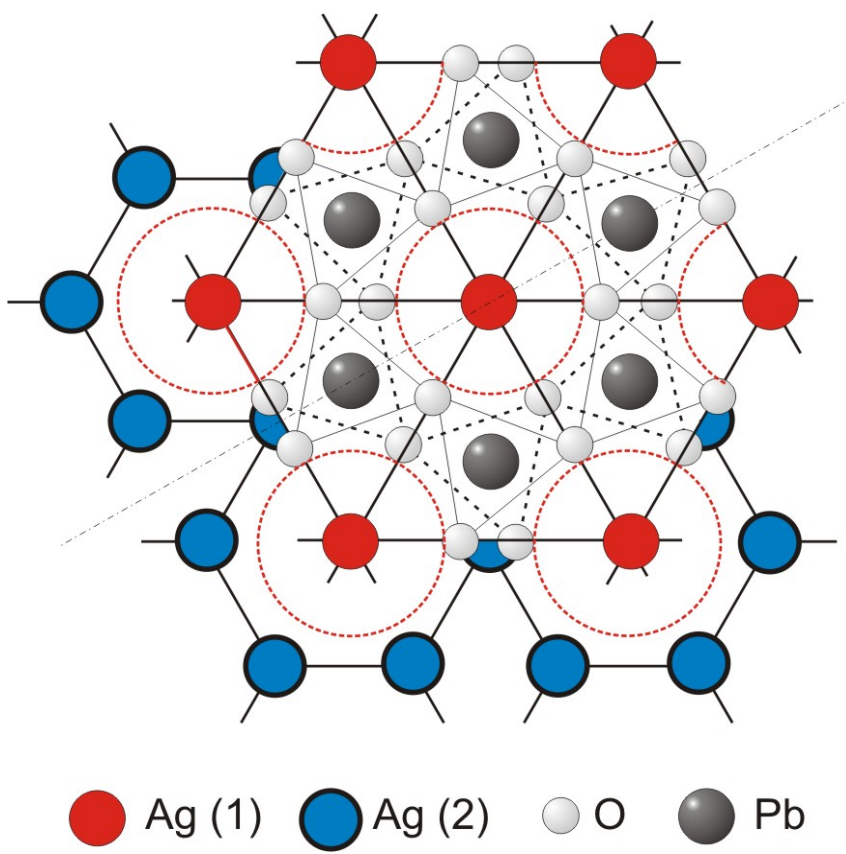

**Figure 2**

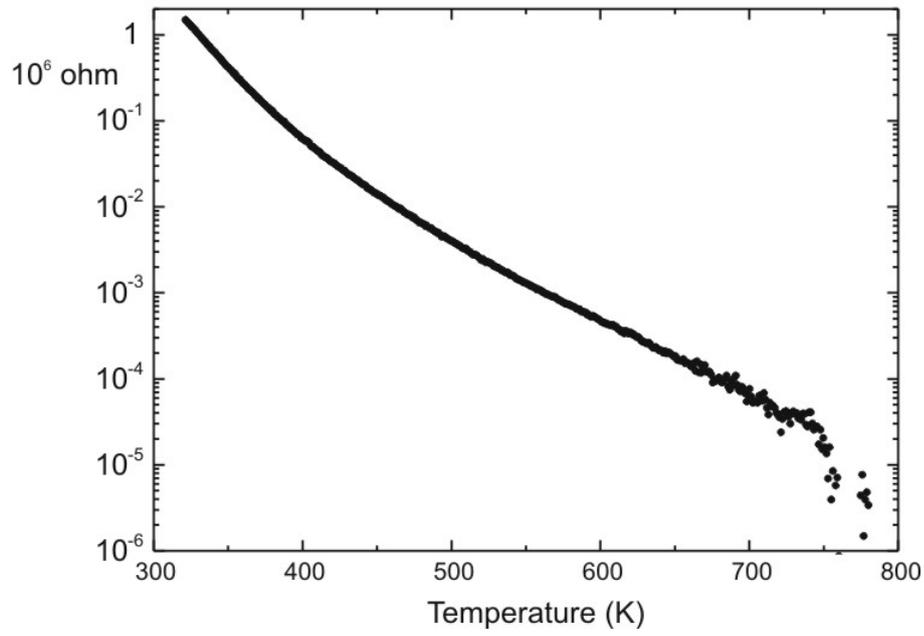

**Figure 3**

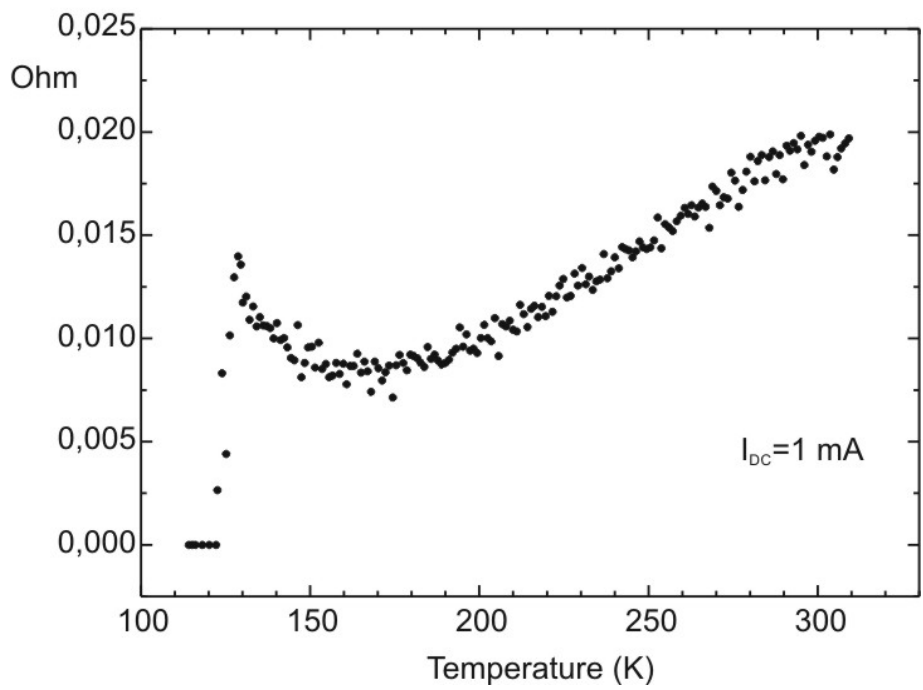

**Figure 4**

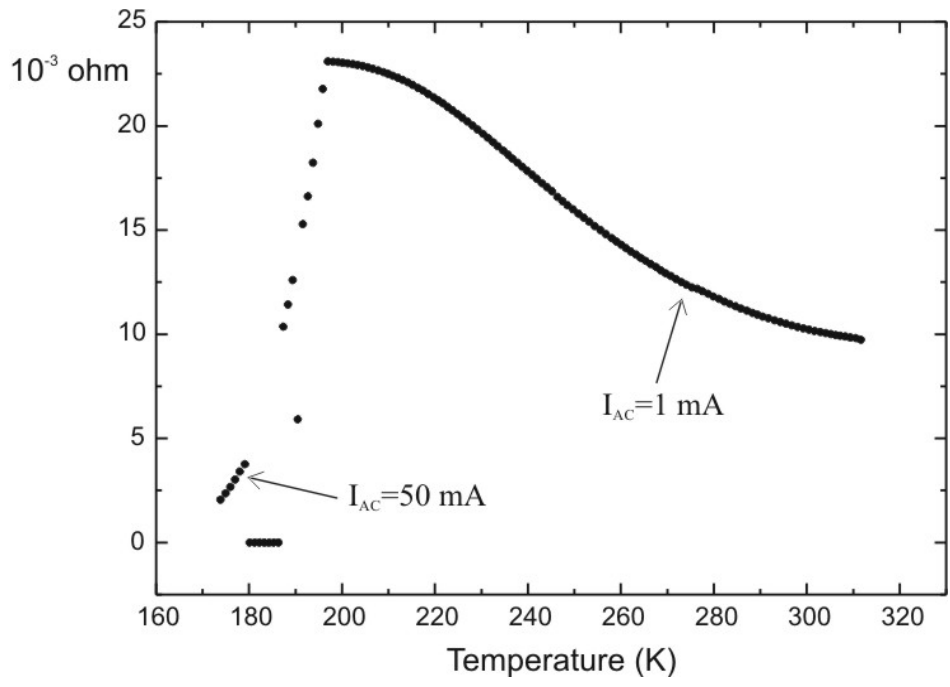

**Figure 5**